# Nonreciprocal nano-optics with spin-waves in synthetic antiferromagnets

Edoardo Albisetti[1,2,*], Silvia Tacchi[3], Raffaele Silvani[3,4], Giuseppe Scaramuzzi[1], Simone Finizio[5], Sebastian Wintz[5], Jörg Raabe[5], Giovanni Carlotti[3], Riccardo Bertacco[1], Elisa Riedo[6,2,*] and Daniela Petti[1,*]

[1]Dipartimento di Fisica, Politecnico di Milano, 20133 Milano, Italy.

[2]Advanced Science Research Center, CUNY Graduate Center, 85 St. Nicholas Terrace, New York, New York 10031, USA.

[3]Istituto Officina dei Materiali del CNR (CNR-IOM), Unità di Perugia, c/o Dipartimento di Fisica e Geologia, Perugia, Italy.

[4]Dipartimento di Fisica e Geologia, Università di Perugia, Via A. Pascoli, Perugia, I-06123, Italy

[5]Paul Scherrer Institute, 5232 Villigen PSI, Switzerland

[6]Tandon School of Engineering, New York University, New York NY 11201, USA

*Correspondence to: edoardo.albisetti@polimi.it, elisa.riedo@nyu.edu, daniela.petti@polimi.it;

## Abstract

**Integrated optically-inspired wave-based processing is envisioned to outperform digital architectures in specific tasks, such as image processing and speech recognition. In this view, spin-waves represent a promising route due to their nanoscale wavelength in the GHz frequency range and rich phenomenology. Here, we realize a versatile optically-inspired platform using spin-waves, demonstrating the wavefront engineering, focusing, and robust interference of spin-waves with nanoscale wavelength. In particular, we use magnonic nanoantennas based on tailored spin-textures for launching spatially shaped coherent wavefronts, diffraction-limited spin-wave beams, and generating robust multi-beam interference patterns, which spatially extend for several times the spin-wave wavelength. Furthermore, we show that intriguing features, such as resilience to back-reflection, naturally arise from the spin-wave nonreciprocity in synthetic antiferromagnets, preserving the high quality of the interference patterns from spurious counterpropagating modes. This work represents a fundamental step towards the realization of nanoscale optically-inspired devices based on spin-waves.**

## Introduction

Spin-waves, or magnons, are wave-like propagating perturbations in the spin-arrangement of magnetic materials.[1] Due to their peculiar properties, such as nanometric wavelength in the GHz-THz frequency range and absence of Joule losses, they represent a promising route for integrated low-power analog and digital computation, and signal processing.[2–5] In this context, surface acoustic waves (SAW) and bulk (BAW) acoustic waves have represented reference technologies for wave-based wireless communication, but their scalability is limited.[6] On the other hand, spin-waves offer the potential for nanoscale integrability and provide an interesting physical system for developing unconventional computational frameworks, such as neural networks and reservoir computing.[7]

To exploit the rich phenomenology of spin-waves[8–14] for integrated optically-inspired processing, generating coherent spatially engineered wavefronts and controlling the propagation and interference of multiple spin-wave beams is crucial. One of the most versatile methods for spin-wave emission is based on using patterned shaped microantennas, for generating a localized oscillating magnetic field in correspondence of the magnetic material.[15,16] However, the emission of spin-waves with sub-micrometric wavelength is challenging, due to the slow spatial decay of the field and the high impedance of the antennas as the size approaches nanoscale dimensions.[17] Ferromagnetic nanostructures,[18,19] spin-transfer torque,[20–22] spin-orbit torques,[23–25] spin-currents,[26] magnetoelastic coupling[27] and multiferroic heterostructures,[28] are alternative methods used for spin-wave generation, however they do not provide a straightforward control of the wavefront and beam shape. Other methods for spatially controlling the spin-wave propagation include using the natural anisotropy of the spin-wave dispersion,[29–32] or physically micro/nanopatterning the spin-wave medium[33,34], but their flexibility and scalability to multi-beam configurations is limited.

Recently, nanoscale spin-textures in magnetic materials have gained attention for their potential as functional elements in spin-wave devices. In particular, spin-wave guiding,[35–40] generation,[41–46] and tunable transmission[47,48] was observed at naturally occurring spin-textures such as vortices and domain walls.

Here, we experimentally realize a nanoscale optically-inspired spin-wave platform, by harnessing the potential of spin-texture nanoengineering to generate and control spatially shaped nanoscale spin-wave wavefronts. This is done by coupling radiofrequency magnetic fields with engineered magnonic nanoantennas, consisting of nanoscale spin-textures. First, we demonstrate the use of thermally assisted magnetic scanning probe lithography (tam-SPL)[49–52] for nanopatterning spin-textures in synthetic antiferromagnets. Then, we show the generation of spin-waves with planar, radial, convex and concave wavefronts, the directional emission of spin-wave beams, and their diffraction-limited focusing into dimensions comparable to their nanoscale wavelength. Furthermore, by combining the emission of multiple nanoantennas, we generate robust interference patterns, which span for more than 15 times the spin-wave wavelength. Finally, we show that the nonreciprocity of spin-waves in our SAF system [53] leads to resilience from back-scattering at defects and spurious reflections, therefore preserving an extremely high quality of the interference patterns. The demonstration of such a rich phenomenology opens up intriguing possibilities for the realization of integrated analog processing nanodevices using spin-waves.

## Discussion

In Figure 1, we present the concept of magnonic nanoantennas, based on nanopatterned shaped magnetic domain walls. The multilayer structure, grown by magnetron sputtering (see Methods) is sketched in Fig. 1a. It consists of two 45 nm CoFeB ferromagnetic layers aligned antiferromagnetically via bilinear coupling through a thin non-magnetic Ru 0.6 nm spacer (synthetic antiferromagnet, SAF). The magnetization direction of the top CoFeB layer is set via exchange bias with a 10 nm IrMn antiferromagnetic layer. The hysteresis loop, shown in the Supplementary Information (Fig. S1), features two well defined and separate lobes, related to the two ferromagnetic layers, and the characteristic flat region around zero external field, indicating robust antiferromagnetic coupling.

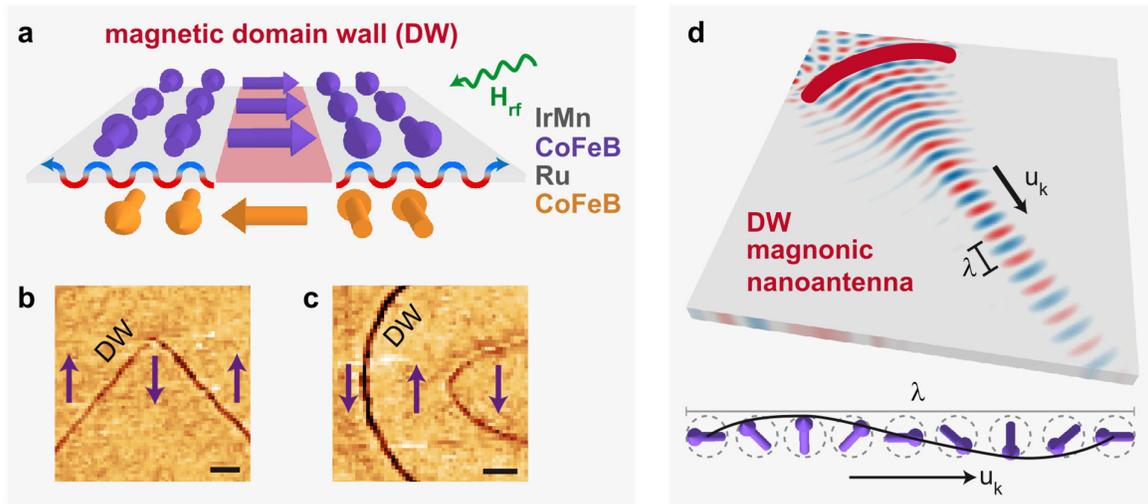

**Figure 1. Magnonic nanoantennas based on patterned spin-textures. a**, Sample structure of the exchange biased synthetic antiferromagnet (SAF), consisting of a sputtered IrMn / CoFeB / Ru / CoFeB magnetic multilayer. The orientation of the spins of a magnetic domain wall (DW) is sketched. Spin-waves are launched by driving the oscillation of the domain wall with a radiofrequency (RF) external magnetic field $H_{rf}$. **b**, **c**, Magnetic force microscopy images of patterned spin-textures. Straight, angled (**b**) and curved (**c**) domain walls with precisely controlled geometry and spin configuration are written in the spin-texture via thermally assisted magnetic scanning probe lithography (tam-SPL). Purple arrows indicate the direction of the equilibrium magnetization in the top CoFeB layer. Scale bars: 3 μm. **d**, Simulation of spatially shaped spin-wave wavefronts generated by a curved nanoantenna. $u_k$ marks the spin-wave propagation direction. Below, the precession of the spins along a single spin-wave wavelength $\lambda$.

For writing the spin-texture of the SAF, a highly localized field cooling is performed via thermally assisted magnetic scanning probe lithography[49] (tam-SPL), by sweeping a heated scanning probe[54–57] on the surface of the sample in an applied external field (see Methods). This allows us to spatially control the exchange bias direction, and therefore the magnetization orientation of the top CoFeB layer, at remanence. Importantly, the magnetization of the bottom CoFeB layer is oriented point-by-point antiferromagnetically due to the interlayer exchange coupling. As a result, two-dimensional magnetic domains, mono-dimensional domain walls and zero-dimensional topological magnetic quasiparticles with deterministically tailored spin configuration are directly patterned in the continuous synthetic antiferromagnet multilayer. Fig. 1a shows the sketch of a straight 180° domain wall (see also Fig. S2 and related discussion in the Supplementary Information). Fig. 1b, c shows magnetic force microscopy (MFM) images of straight (b) and curved (c) magnetic domain walls, patterned via tam-SPL by writing domains with oppositely oriented magnetization direction. In each image, the equilibrium magnetization direction of the top CoFeB layer is indicated by violet arrows. In the following, we show that spin-textures can be used as magnonic nanoantennas, for spatially shaping at the nanoscale and manipulating spin-wave wavefronts propagating in the film. Panel D shows a micromagnetic simulation sketching this concept, where a curved domain wall, indicated by the thick magenta line, driven into oscillation by a microwave field, directionally emits spin-waves with wavelength λ along the direction $\mathbf{u_k}$. Below, the orientation of the spins in the top CoFeB layer within a single spin-wave wavelength is sketched.

In order to show the versatility and potential of this platform, we provide the proof-of-concept of different functionalities. In Figure 2, we demonstrate the generation of spatially shaped wavefronts and directional emission of spin-wave beams by using curved nanoantennas.
The spin-wave excitation and propagation is visualized stroboscopically with high spatial and temporal resolution via Scanning Transmission X-Ray Microscopy (STXM) (see Methods). The black/white color corresponds to the oscillation of the out-of-plane component of the magnetization $M_z$ associated to the propagation of spin-waves, with respect to the average value over one period of oscillation. The oscillation of the domain wall is driven by an external magnetic field $H_{RF}$, provided by a stripline run by radio-frequency current, patterned on top of the multilayer in close proximity to the wall. Specifically, in Supplementary Figure S3 and related discussion, we show that the wall dynamics is mainly excited by the coupling of the oscillating out-of-plane component of $H_{RF}$ with the out-of-plane component of the magnetization in proximity of the wall. This mechanism has important advantages compared to in-plane fields. In fact, in this case, the SW excitation efficiency does not depend on the relative orientation of the stripline and the spin-textures, therefore allowing spin-wave emission from curved and angled sections of the domain walls, consistently with the experimental results shown in Figure 2.
Panels a and b show the experimental STXM image (left, see also Supplementary Movie 1) and micromagnetic simulation (right) of the emission of spin-waves from an extended curved domain wall, indicated by the thick magenta line. Subsequent convex wavefronts are indicated by alternated red/blue lines, and white arrows indicate the direction of the equilibrium magnetization in the top layer. Noteworthy, the wavefronts retain the shape of the emitter, so that a straightforward wavefront engineering is enabled, by tuning the curvature of the domain wall.

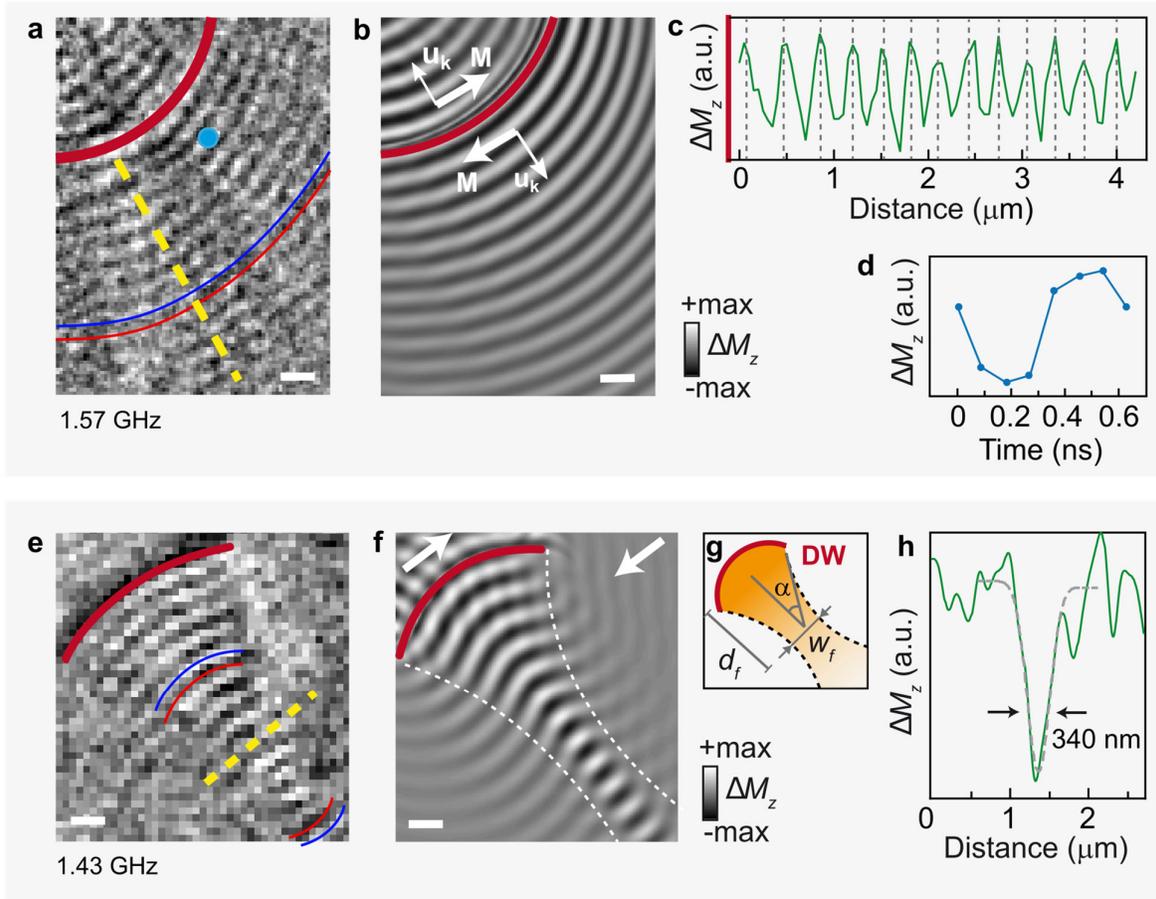

**Figure 2. Spin-wave wavefront engineering. a**, **b,** Experimental STXM image (**a**) and micromagnetic simulation (**b**) of the directional emission of convex spin-wave wavefronts by a curved domain wall (red line). Wavefronts are indicated by thin red and blue lines. **c, d,** Spatial and temporal profiles extracted from the yellow dashed line and blue dot in panel **a**, and corresponding Supplementary Movie 1, respectively. Strong spin-wave intensity is measured after more than 15 periods of propagation. **e, f,** Experimental image (**e**) and simulation (**f**) of the emission and focusing of spin-wave beams with concave wavefronts. **g,** Diffractive-optics analogue of spin-wave focusing by a magnonic nanoantenna with angular aperture 2α. **h,** Spatial profile of the spin-wave amplitude along the dashed line in panel **e**. At the beam waist, located 2.5 µm away from the emitter, spin-waves are localized in a region comparable to the spin-wave wavelength ~ 340 nm. White arrows indicate the direction of the equilibrium magnetization M in the top CoFeB film, and the spin-wave wavevector direction $u_k$. Scale bars: 500 nm.

Panel c, d show the experimental spatial and temporal profiles extracted from the dashed line and blue dot in panel a, respectively, featuring a spin-wave wavelength of λ ~ 300 nm for an excitation frequency $f$ = 1.57 GHz. Importantly, a strong spin-wave signal is clearly visible after more than 15 periods of propagation, limited by the STXM acquisition window. The combination of short-wavelength spin-wave modes, wavefront engineering, and propagation length exceeding several times their wavelength is essential for building analog nanodevices based on spin-wave interference. Another appealing feature in wave-based processing is the possibility of focusing the wave energy in specific and controlled directions. In panels e, f, we demonstrate the directional emission and focusing of spin-wave beams. The left panel shows the STXM image (see also Supplementary Movie 2) of propagating spin-waves emitted by the curved portion of a domain wall nanoantenna, in magenta. The wavefront shape is highlighted in red (trough) and blue (crest). The right panel shows the corresponding micromagnetic simulation, and the direction of the equilibrium magnetization in the top layer (white arrows). The wall locally excites a converging spin-wave beam which is focused at a distance $d_f \sim 2.5$ µm from the emitter. Consistently with a focusing effect, the concavity of the wavefront changes from concave to convex after the focal point. Panel H shows the spatial profile extracted in correspondence of the yellow dashed line in panel E, at the focal point. The full width at half maximum (FWHM) of the beam amplitude at the focal point is $w_f \sim 340 \pm 50$ nm. Noteworthy, this picture is in agreement with a description analogue to diffractive optics for the emission from a source of finite size, as shown in panel G. The focusing is determined by the numerical aperture of the object, in our case the finite wall source, so that the minimum beam width can be estimated as $w_f = \lambda/[2\sin(\alpha)] = 363$ nm, where $\lambda = 330$ nm is the experimental spin-wave wavelength at $f = 1.43$ GHz and $\alpha = 27°$ is determined by the wall geometry.

The generation of controlled wavefronts is one of the crucial requirements for analog processing. In the following, we show that by combining multiple spin-wave sources, we are able to generate robust spin-wave interference patterns which span for several times their wavelength. For demonstrating the generality of the approach, we use planar wavefronts emitted by straight domain walls, and radial wavefronts emitted by vortices (see Supplementary Note 6 and Supplementary Movies 3, 4) as building blocks for generating interference patterns.

In Figure 3a (experiment, see also Supplementary Movie 5) and 3b (simulations), a straight domain wall (magenta line) and a vortex located within the domain wall itself (magenta circle) are used for generating simultaneously linear and radial wavefronts, respectively, which spatially superimpose during propagation. The curved and straight red/blue lines identify the separate wavefronts with radial and linear symmetry, respectively. In the STXM images and simulations, the black/white contrast represents $\Delta M_z$ component at a specific time. Destructive interference fringes, where the spin-waves sum in anti-phase are visible as low intensity, grey regions, while constructive interference is visualized as high intensity black/white regions. For better visualizing the interference pattern, in panel c we show the spatial map of the spin-wave amplitude, obtained by evaluating point-by-point the amplitude of the experimental oscillation of $\Delta M_z$. The characteristic alternated interference minima and maxima are visible as yellow and blue regions, respectively, and are numbered in the figure. Panel d shows the experimental spatial profile extracted from the red dashed line in panel C (red line), showing the first 5 interference maxima spaced by $\sim 450$ nm, in excellent agreement with the simulation (gray line). In panels e (experiment, see also Supplementary Movie 6), f (simulations), a domain wall comprising two straight branches forming a $\beta = 135°$ angle with each other is used for directionally emitting two angled wavefronts with planar symmetry, indicated by the red/blue lines. The emission of the planar wavefronts and the interference figure are clearly visible both in the STXM image and simulation. Noteworthy, the interference patterns are still clearly visible after more than 15 oscillation periods, limited by the experimental acquisition window. These results demonstrate the generation of robust multi-beam interference patterns at the nanoscale, i.e. the basic element for integrated spin-wave based processing.

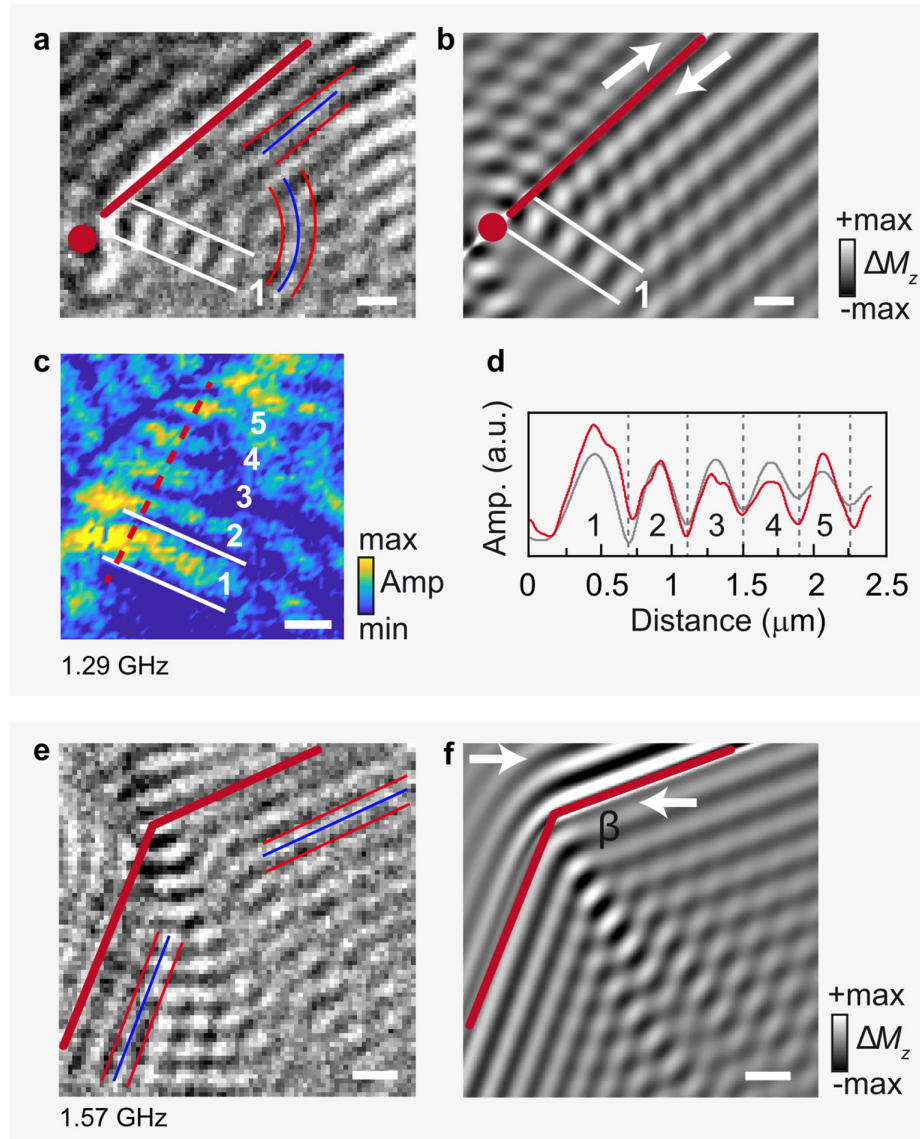

**Figure 3. Generation of multi-beam interference patterns. a, b** Experimental STXM image (**a**) and simulations (**b**) of the interference from radial wavefronts emitted by a vortex and planar wavefronts emitted from a straight domain wall. The domain wall and vortex are indicated by magenta lines and dot. **c**, Spatial map of the spin-wave amplitude related to panel **a**. Constructive and destructive interference fringes are visible

as alternated minima (blue) and maxima (yellow). The region of the first interference maximum is indicated by white lines in panels **a-c**. **d,** Experimental (red) spatial profile of the spin-wave amplitude extracted from the red dashed line in panel **c**, and corresponding simulation (grey). The numbers refer to the first 5 interference maxima. **e, f,** Experimental (**e**) and simulated (**f**) interference pattern generated by the spin-wave wavefronts emitted by two angled domain wall nanoantennas, indicated in magenta. The two linear wavefronts are indicated by thin red and blue lines. White arrows indicate the equilibrium magnetization direction of the top layer. Scale bars: 500 nm.

The high quality and robustness of the interference patterns are determined by the peculiar properties of the spin-wave dispersion in SAFs. In Figure 4, we show the results of micromagnetic simulations of the spin-wave propagation in extended exchange biased antiferromagnetically coupled bilayers (see methods and Supplementary Note 7), focusing on the in-plane angular dependence of the dispersion. In Figure 4a, the spin-wave dispersion curves calculated at remanence for different values of the angle $\varphi$ between the magnetization of the top CoFeB film ($M_{top}$) and the wavevector $k$ are shown. Black filled dots show the experimental dispersion of spin-waves emitted by a straight domain wall, confirming the possibility of tuning the spin-wave wavelength by varying the microwave excitation frequency. Two fundamental features are evident from the dispersion.

First, the strong frequency nonreciprocity causes the two branches ($\pm k$) to have significantly different dispersion for $\varphi$ larger than 15° (see also Supplementary Note 7). In fact, the two branches have different group velocity ($\partial\omega/\partial k$) and wavelength, for a fixed frequency. Noteworthy, below the threshold frequency value corresponding to the crossing point of the different curves at $k = 0$ ($f = 1.57$ GHz), spin-wave propagation occurs only with positive wavevector ($+k$). In our experiments, we focus on the short-wavelength spin-waves belonging to this branch. For $\varphi$ below 15°, the nonreciprocity is lost, as shown by the red curve in the dispersion, corresponding to $\varphi = 0°$ where the spin-wave wavevector is parallel to the magnetization (backward (BA) configuration).

Second, such $+k$ short-wavelength modes are characterized by a low-anisotropy on a wide angular range, i.e. their wavelength and group velocity vary weakly with the propagation direction in a wide range of angles (from 90°, corresponding to the "Damon-Eshbach" (DE) configuration, dark blue curve, down to 15°, orange curve).

The effect of the combination of these two properties is evident in Fig. 4b, where we show the simulated one-way propagation of wavefronts emitted by a point source (see also Supplementary Note 5) sinusoidally oscillating at 1.5 GHz. One can see that spin-waves, with highly regular radial wavefronts, are emitted only towards the right half-plane from the source, reflecting the above discussed nonreciprocity. A deeper understanding can be achieved by analyzing the calculated isofrequency curves in the reciprocal space (also known as slowness curves), plotted in Fig. 4c. For negative $k_x$ the isofrequency curves are characterized by a strongly anisotropic behavior, typical of dipole-dominated spin-waves. On the contrary, for positive $k_x$, the contours show an elliptical shape reflecting the low-anisotropy character of the short wavelength modes.

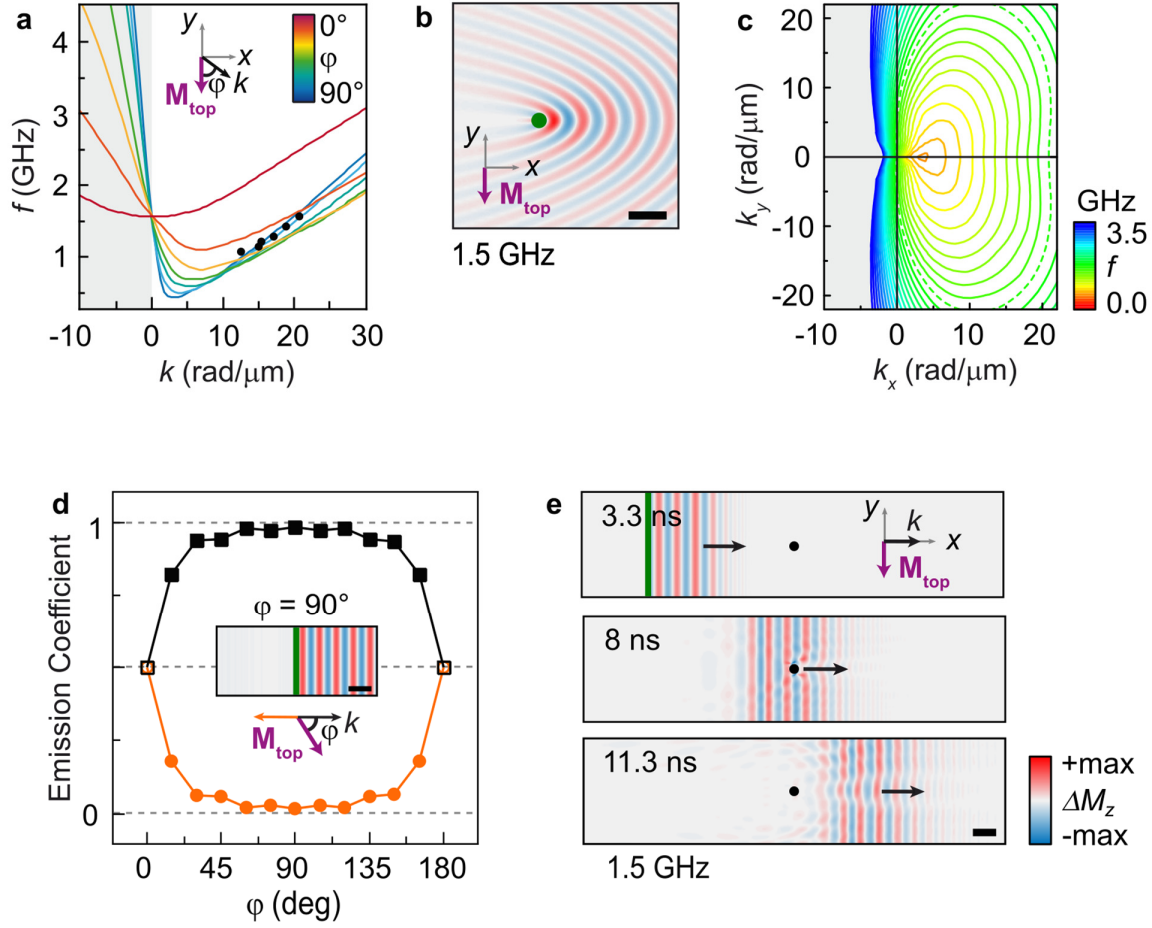

**Figure 4. Characteristics of nonreciprocal spin-wave modes in SAF. a**, Color-coded lines show the simulated spin-wave dispersion in the SAF multilayer, as a function of the angle $\varphi$ between the wavevector $k$ and the equilibrium magnetization $M_{\text{top}}$ in the top layer. Black dots show the experimental points. On the right side (positive wavevectors), the dispersion features short-wavelength spin-waves characterized by low anisotropy. **b**, Simulated spin-wave wavefronts, in the real space, emitted by a point excitation (green dot) at $f$ = 1.5 GHz. The direction of the magnetization of the top layer $M_{\text{top}}$ is indicated by the purple arrow. Scale bar: 500 nm. **c**, Simulated isofrequency curves, in the reciprocal space, for the spin-wave modes propagating in the SAF. The equilibrium magnetization of the top layer $M_{\text{top}}$ is directed along -$y$. The dashed line indicates the curve for $f$ = 1.5 GHz. **d**, Simulated nonreciprocal emission coefficient as a function of $\varphi$ for spin-waves emitted on the right (black curve and squares) and on the left (orange curve and circles) of the excitation region (green line). The inset shows the spin-wave wavefronts in the case of $\varphi = 90°$. **e**, Snapshots at different times of simulated spin-wave packets excited for $\varphi = 90°$, before and after reaching a 200 nm wide non-magnetic circular defect.

To gain more insight on the nonreciprocity of the spin-wave amplitude, the spin-wave emission has been investigated by micromagnetic simulations for different orientations of the magnetization ($\varphi$ angle). Fig. 4d shows the nonreciprocal emission coefficient, i.e. the spin-wave amplitude on the right (dark line and squares) and left (red line and circles) side of the line excitation region, normalized to the total amplitude. The system is excited at a frequency of 1.5 GHz. It can be seen that the emission on the right side of the simulated area amounts to above 94% of the total emission for a wide angular range of about ±70° around the DE configuration ($\varphi = 90°$). This indicates that the spin-wave emission is strongly nonreciprocal for a wide range of orientations of the magnetization. This amplitude nonreciprocity is lost when $\varphi$ approaches 0° or 180°. In these cases, simulated at an excitation frequency of 2 GHz, the emission is reciprocal and the emission coefficient tends to the limiting value of 0.5.

We note that, in our experiments, emission and propagation of spin-waves with the same wavelength and amplitude is observed on both sides of the domain wall (see e.g. Figure 2a). This is perfectly consistent with the nonreciprocity of the system, since the relative orientation between the wavevector $k$ and the magnetization $M$ is the same at the two sides of the wall (see also Supplementary Note 8). The peculiar combination of low anisotropy and nonreciprocal propagation is at the origin of the features we observe experimentally, i.e. highly regular wavefronts, which are maintained for distances of several times the wavelength, and the absence of back-reflections by boundaries or defects, which preserve the spin-wave interference patterns from the contamination of counter-propagating modes. In order to further demonstrate the resilience to back-reflection, the propagation of spin-wave packets in presence of a non-magnetic defect has been investigated by micromagnetic simulations. A defect, consisting of a 200 nm wide cylindrical hole is defined in the whole thickness of the SAF at a distance of 2.5 μm from the excitation region. The system is excited by a line source at a frequency of 1.5 GHz. Fig. 4e shows three snapshots of spin-wave packets excited in the DE configuration ($\varphi = 90°$), before and after reaching the defect. It is observed that the nonreciprocity in SAF ensures that spin-waves impinge on the defect without sizeable back-reflection. Noteworthy, such robustness with respect to back-reflection is not observed in single magnetic layers in the DE configuration (as illustrated in Supplementary Note 9).

We highlight that the nanoscale control of the spin configuration, e.g. via tam-SPL, allows to exploit nonreciprocity as an additional design rule for the control of the emission and propagation of spin-waves.

## Conclusion

In this work, we realized a versatile optically-inspired platform for controlling the generation, propagation and interference of short-wavelength spin-waves. We demonstrated the spatial engineering of spin-wave wavefronts via tailored magnonic nanoantennas, the focusing of spin-wave beams down to the nanoscale and the controlled generation of robust multibeam interference patterns which span multiple times the spin-wave wavelength. Furthermore, we showed that controlling spin-waves in synthetic antiferromagnets offers the unique opportunity to combine concepts borrowed from optics, with phenomena naturally arising from the nonreciprocal spin-wave dispersion, such as resilience to back-reflection and defects. In the route towards applications, we point out that this platform is based on CMOS compatible magnetic thin films grown via sputtering, does not require bulky electromagnets for applying external fields, and allows for the full reconfigurability of the magnonic nanoantennas, e.g. via laser writing or t-SPL.

The unique combination of these features gives rise to a versatile playground for studying the physics of nonreciprocal spin-wave propagation, and represents a fundamental step towards optically-inspired spin-wave processing. Future applications include reconfigurable microwave filters, isolators, and devices for pattern and speech recognition based on the emission, manipulation and interference of coherent spin-wave wavefronts at the nanoscale.

**Methods**

**Sample fabrication**

$Co_{40}Fe_{40}B_{20}$ 45 nm / Ru 0.6 nm / $Co_{40}Fe_{40}B_{20}$ 45 nm / $Ir_{22}Mn_{78}$ 10 nm / Ru 2 nm stacks were deposited on 200 nm thick $Si_3N_4$ membranes by DC magnetron sputtering using an AJA Orion8 system with a base pressure below $1\times10^{-8}$ Torr. During the deposition, a 30 mT magnetic field was applied in the sample plane for setting the exchange bias direction in the as-grown sample. Then, the samples underwent an annealing in vacuum at 250 °C for 5 minutes, in a 400 mT magnetic field oriented in the same direction as the field applied during the growth.

Thermally assisted magnetic Scanning Probe Lithography (tam-SPL) was performed via NanoFrazor Explore (SwissLitho AG). Spin-textures were patterned by sweeping in a raster-scan fashion the scanning probe, heated above the blocking temperature of the exchange bias system $T_B \approx 300°$ C, in presence of an external magnetic field. Two rotatable permanent magnets were employed for generating a uniform external magnetic field applied in the sample plane during patterning.

Microstrip antennas were then fabricated via optical lithography using a Heidelberg MLA100 Maskless Aligner and lift-off, after depositing a 50 nm thick $SiO_2$ insulating layer via magnetron sputtering. A Cr 7 nm / Cu 200 nm bilayer was deposited by means of thermal evaporation with an Evatec Bak 640 system.

**Scanning trasmission X-ray microscopy**

The magnetic configuration of the samples was investigated with time-resolved scanning transmission X-ray microscopy at the PolLux (X07DA) endstation of the Swiss Light Source[58]. In this technique, monochromatic X-rays, tuned to the Co L3 absorption edge (photon energy of about 781 eV), are focused using an Au Fresnel zone plate with an outermost zone width of 25 nm onto a spot on the sample, and the transmitted photons are recorded using an avalanche photodiode as detector. To form an image, the sample is scanned using a piezoelectric stage, and the transmitted X-ray intensity is recorded for each pixel in the image. The typical images we employed for the investigation of the spin-wave propagation in our samples were acquired with a point resolution between 50 nm and 100 nm.

Magnetic contrast in the images is achieved through the X-ray magnetic circular dichroism (XMCD) effect, by illuminating the sample with circularly polarized X-rays. As the XMCD effect probes the component of the magnetization parallel to the wave vector of the circularly polarized X-rays, the samples were mounted to achieve perpendicular orientation of the surface with respect to the X-ray beam, allowing us to probe the out-of-plane component of the magnetization in the SAF.

The time-resolved images were acquired in a pump-probe scheme, using an RF magnetic field, generated by injecting an RF current in a microstrip antenna as pumping signal and the X-ray flashes generated by the synchrotron light source as probing signal. The pumping signal was synchronized to the 500 MHz master clock of the synchrotron light source (i.e. to the X-ray flashes generated by the light source) through a field programmable gate array (FPGA) setup. Due to the specific requirements of the FPGA-based pump-probe setup installed at the PolLux endstation, RF frequencies of $f_{\text{exc}} = 500 \times M/N$ [MHz], being $N$ a prime number and $M$ a positive integer, were accessible. For the measurements presented in this work, $N$ was typically selected to be equal to 7, giving a phase resolution of about 50° in the time-resolved images. Depending on the RF frequency, the temporal resolution of the time-resolved images is given by $2/M$ [ns], with its lower limit given by the width of the X-ray pulses generated by the light source (i.e. about 70 ps FWHM).

For extracting the spatial map of the spin-wave amplitude from the STXM video, the time-trace of each pixel was fitted with a sinusoidal function, whose amplitude was extracted.

**Magnetic Force Microscopy**

Magnetic force microscopy of the patterned magnetic spin-textures was performed via a Bruker Multimode 8 scanning probe characterization system, equipped with a Nanosensors PPP-MFMR AFM magnetic probe. MFM imaging was performed in tapping lift-mode.

**Micromagnetic Simulations**

Micromagnetic simulations of the magnetization dynamics were carried out by solving the Landau–Lifshitz–Gilbert equation of motion, using the open-source, GPU-accelerated software MuMax[3]. The simulated material parameters were set to the following values: saturation magnetization $M_s$=1000 kA·m$^{-1}$, exchange constant $A_{ex}$=1.2·10$^{-11}$ Jm$^{-1}$, interlayer exchange coupling constants J= -0.6 mJ·m$^{-2}$. The exchange bias field was modeled applying in the uppermost layer of cells of the top CoFeB film an external magnetic field of 30 mT parallel to the magnetization of the CoFeB film itself. Periodic boundary conditions in the *y* direction were used (see Supplementary Note 1 for further details).

**Acknowledgements**

The authors thank A. Melloni, F. Morichetti, P. Laporta, P. Biagioni for fruitful discussions and Guido Gentili for the simulations of the electrical behavior of the striplines. The research leading to these results has received funding from the European Union's Horizon 2020 research and innovation programme under grant agreements no. 705326, project SWING, and no. 730872, project CALIPSOplus. This work was partially performed at Polifab, the micro- and nano-technology center of the Politecnico di Milano. Part of this work was performed at the PolLux (X07DA) endstation of the Swiss Light Source, Paul Scherrer Institut, Villigen, Switzerland. The PolLux endstation was financed by the German Minister für Bildung und Forschung (BMBF) through contracts 05KS4WE1/6 and 05KS7WE1.

# Nonreciprocal nano-optics with spin-waves in synthetic antiferromagnets

## SUPPLEMENTARY INFORMATION

### Supplementary Note 1. Micromagnetic simulations

To simulate the spatial profile of the spin-wave modes emitted by the spin-textures, the total simulated volume was discretized into cells having dimensions of $10 \times 10 \times 11.25$ nm$^3$. First, the system was relaxed to the ground state for stabilizing the spin-texture and then the magnetization dynamics was excited applying a 0.1 mT time-varying sinusoidal magnetic field to the whole system. The Gilbert damping parameter was set to $\alpha = 0.008$.

To compute the spin wave dispersion relation in the extended antiferromagnetically coupled bilayer, a stripe geometry with a length of 12.6 µm (along $x$), a width of 40 nm (along $y$) and a total thickness of 2×45 nm for the two CoFeB films was modeled. The total simulated volume was discretized into 2048×4×16 number of cells and periodic boundary conditions were applied in the x and y directions. To simulate the spin wave dispersion for different configurations, the magnetization of the two CoFeB films was initialized to be pointing in opposite direction at remanence, following the relation $M_s(\cos(\theta + \pi), \sin(\theta + \pi), 0)$ $[M_s(\cos \theta, \sin \theta, 0)]$ in the top (bottom) one, where $\theta$ is the angle between the $x$-axis and the magnetization of the bottom film (where in Fig. 4A the angle $\varphi = \pi - \theta$). In order to excite spin-waves, a sinc-shaped field pulse $b(t) = b_0 \frac{\sin(2\pi f_0(t-t_0))}{2\pi f_0(t-t_0)}$, directed along the z-axis, with amplitude $b_0 = 10$ mT and frequency $f_0 = 30$ GHz, was applied in the center of the simulated area in a region having a size of 6 nm and 40 nm along the $x$ and $y$ directions, respectively. The dispersion relation was calculated by performing a Fourier-transform of the $z$-component of the magnetization both in space and time in the whole simulated area.

To compute the spin wave nonreciprocal emission coefficient in the extended antiferromagnetically coupled bilayer (Fig. 4D), a stripe geometry with a length of 25 µm (along $x$), a width of 40 nm (along $y$) and a total thickness of 2×45 nm for the two CoFeB films was modeled. The total simulated volume was discretized into 4096×4×16 number of cells and periodic boundary conditions were applied in the $x$ and $y$ directions. To simulate the spin-wave dispersion for different configurations, the magnetization of the two CoFeB films was initialized as in the previous micromagnetic simulations. The magnetization dynamics was excited applying a 5 mT time-varying sinusoidal magnetic field to the center of the system, in a region 24 nm wide. The Gilbert damping parameter was set to $\alpha = 0.001$. The non-reciprocal emission coefficient has been calculated as:

$$\text{Emission Coefficient} = \frac{m_i(f_0, x_i)}{m_R(f_0, x_R) + m_L(f_0, x_L)} \quad \text{with } i = R, L$$

Where $m_R$ and $m_L$ is the integrated intensity of the time-Fourier transform of the magnetization dynamic in the cells at a distance of 2 μm and -2 μm, respectively, from the excitation region.

**Supplementary Note 2. Magnetic characterization of exchange biased SAF samples**

Figure S1 (top panel) shows the hysteresis loop of the CoFeB 45 / Ru 0.6 / CoFeB 45 / IrMn 10 nm / Ru 2 nm synthetic antiferromagnet (SAF) measured at room temperature via vibrating sample magnetometry (VSM). The sample measurement was performed after initialization, i.e. after undergoing a field cooling in vacuum in a 400 mT magnetic field oriented in the $+y$ direction within the plane of the film. This process sets a uniform exchange bias direction in the system, which pins the magnetization of the top CoFeB layer (purple arrows) along $+y$.

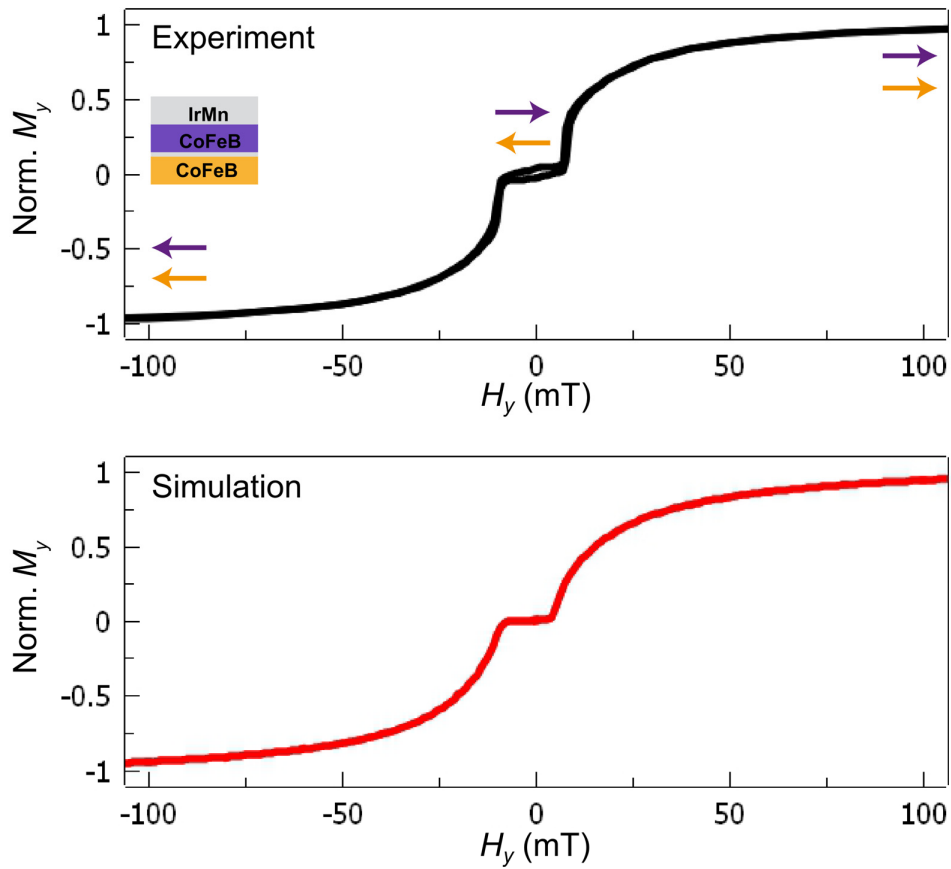

**Figure S1. Experimental and simulated magnetic hysteresis loops of SAF**. The top panel shows the normalized *y*-component of the magnetization of the SAF samples used in the experiments, measured via Vibrating Sample Magnetometry. Purple and orange arrows show the direction of the magnetizations in the top and bottom layer, respectively. The bottom panel shows the corresponding micromagnetic simulation.

The VSM loop shows the *y*-component of the magnetization, normalized to its maximum value, as a function of the external magnetic field applied along *x* in the -100 mT – +100 mT range. Purple and orange arrows show the direction of the magnetization of the top layer and bottom layer, respectively. For strong, negative fields, the magnetization of both layers is saturated in the -*y* direction. By decreasing the field, the RKKY-mediated coupling forces the canting of the magnetization towards antiparallel alignment. At the same time, the exchange bias in the top layer (which is set along +*y*), forces the orientation of the top layer magnetization along +*y*. The combination of these two interactions makes sure that at remanence the magnetizations of the two layers are aligned antiferromagnetically, and that, importantly, the in-plane orientation of the magnetizations is determined by the direction of the exchange bias in the top layer. The robustness of the antiferromagnetic coupling is confirmed by the characteristic plateau observed at low fields. Finally, by applying strong external magnetic field in the +*y* direction, the antiferromagnetic coupling is overcome, and the magnetization of both layers saturate along +*y*.

The bottom panel shows the simulated hysteresis loop of the SAF system, which is in excellent agreement with the experimental one.

## Supplementary Note 3. Spin-configuration of patterned domain walls in SAF

Figure S2 shows static micromagnetic simulations of the spin-configuration of domain walls in SAF. The direction of the magnetization in the two layers is kept antiparallel point-by point by the RKKY interaction, both in the domains and within the domain wall. The right panel shows the direction of the magnetization in the top (purple rectangle) and bottom (orange rectangle) CoFeB layer. The color-code is referred to the out-of-plane component of the magnetization ($M_z$), where positive (negative) $M_z$ is marked by red (blue) color. An increase in the color contrast approaching the central part of the DW is observed, and the change from red to blue contrast across the domain wall. This is consistent with a partial canting of the magnetization out-of-plane in correspondence of the domain wall, which is towards the positive (negative) *z* direction in the left (right) part of the wall. The direction of the out-of-plane canting is determined by the chirality of the wall, i.e. by the direction of the spins at the center of the domain wall. The bottom panel shows the section of the domain wall (*xz* plane), in which the orientation of the magnetization in both layers is visible, together with the color-coded $M_z$. Black arrows indicate the direction of the magnetization in the top (purple) and bottom (orange) layer.

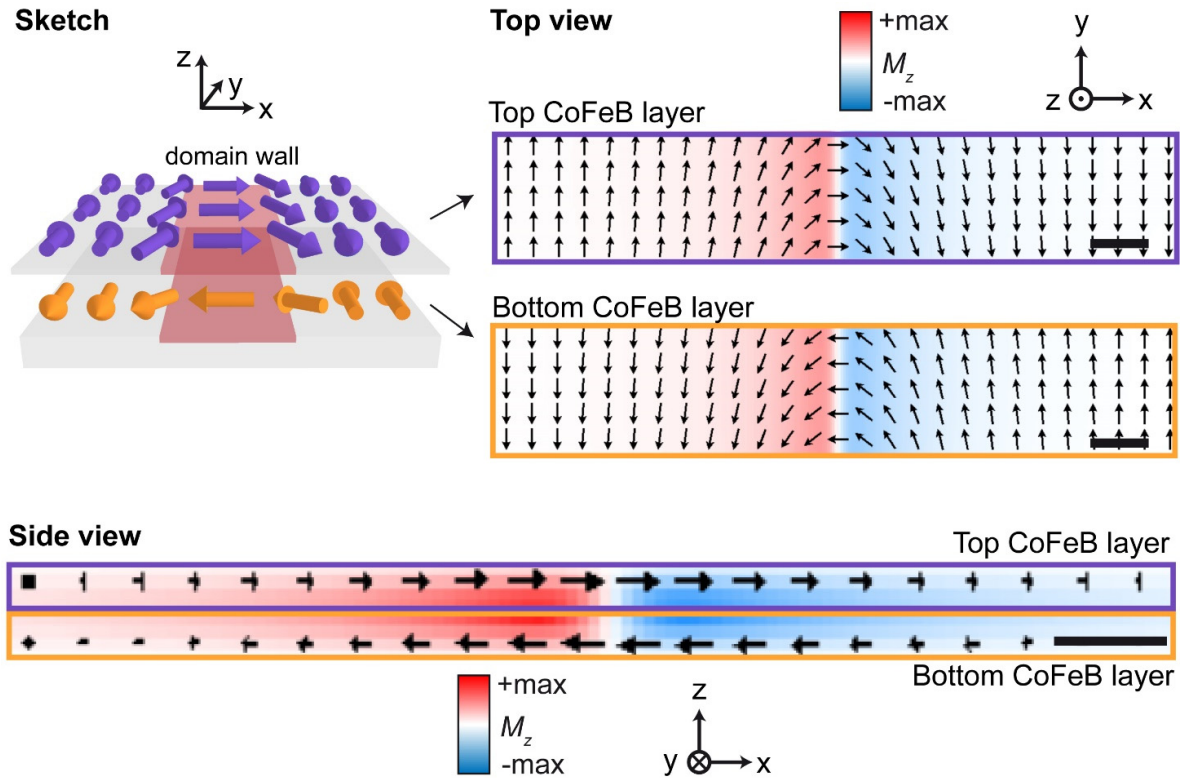

**Figure S2. Spin-configuration of domain walls in SAF**. The left panel shows the sketch of the spin-texture of the domain wall patterned in the exchange biased synthetic antiferromagnet (SAF). The right panel shows the spin-configuration in the top (purple) and bottom (orange) CoFeB layers, respectively. The bottom panel shows the section of the sample, where the top and bottom layers are highlighted by purple and orange rectangles. The black arrows mark the direction of the magnetization, while the color code marks the out-of-plane component of the magnetization ($M_z$), where red (blue) indicates +*z*(-*z*). Scale bars: 200 nm.

**Supplementary Note 4. Coupling of external RF field with nanoscale spin-textures**

Conventionally, spin-waves are excited by the Oersted magnetic field generated by running RF currents in coplanar waveguides or striplines fabricated in proximity of the magnetic medium. The minimum spin-wave wavelength which can be excited in this manner is comparable to the width of the stripline. Due to the slow spatial decay of the magnetic field outside the stripline, and the increase of the impedance of the waveguides when they approach nanoscale dimensions, the generation of nanoscale spin-waves using the direct coupling between nanoscale waveguides and magnetic material is not efficient. Furthermore, the integration of multiple nanoscale waveguides and the spatial shaping of the Oe field profile using curved waveguides is complex and has limitations imposed by the geometry of the contacts.

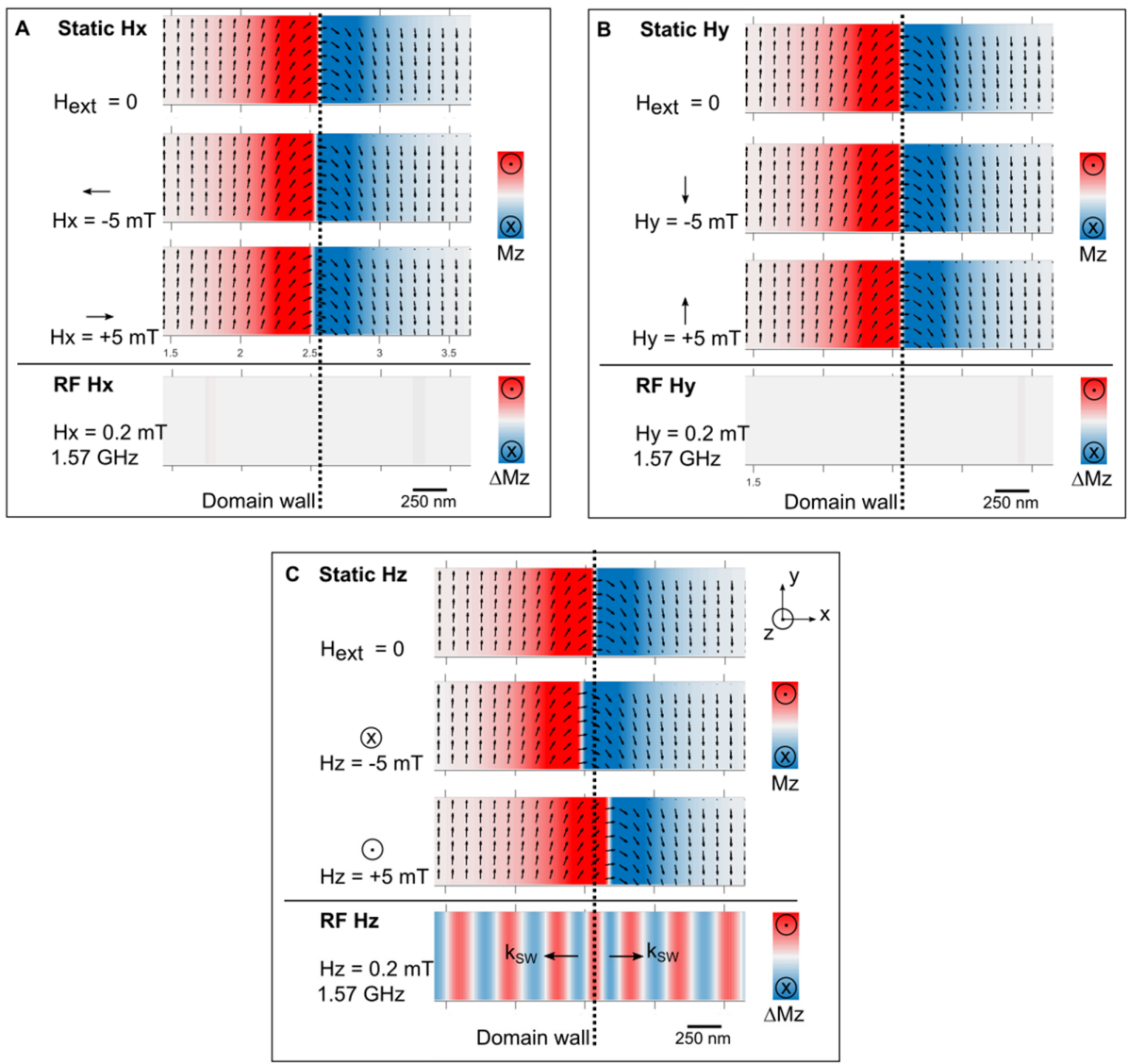

**Figure S3. Coupling of external fields with nanoscale spin-textures.** Panels A, B and C show the effect of a static and oscillating magnetic field with different orientation (along *x*, *y*, and *z*, respectively) on the spin-configuration of a perfectly compensated SAF domain wall. The in-plane spin-configuration is marked by black arrows, and the out-of-plane component is marked by red-blue contrast. No domain wall oscillation and spin-wave emission are observed for in-plane fields (panels A, B). Domain wall oscillation and spin-wave emission is observed when an out-of-plane magnetic field is applied (panel C).

The spin-wave excitation mechanism employed in this work is based on coupling the RF magnetic field generated by a 5 μm stripline with nanoscale spin-textures. In order to investigate the spin-wave emission mechanism, we carried out micromagnetic simulations studying the coupling of static and RF fields with different spatial orientation (both in-plane and out-of-plane) with a domain wall, used for SW excitation. We consider the ideal case of a perfectly compensated SAF structure CoFeB 45 nm / Ru 0.6 nm / CoFeB 45 nm, such as the one used experimentally, where a 180 deg side-to-side domain wall is stabilized in the center of the stripe by imposing an uniaxial anisotropy of intensity $K_u$ = 1x10$^3$ J/m$^3$ along *y*. Note that this is the worst case-scenario regarding the SW excitation efficiency. The orientation of the magnetization within the domains was chosen consistently with our experimental configuration (see e.g. Figure 1).

In Figure S3 we report the spin-configuration of the top SAF layer, in proximity of the wall, following the application of static and RF external field along *x* (panel A), *y* (panel B), and *z* (panel C). In particular, the black arrows mark the in-plane magnetization direction, and the red-blue contrast mark the out-of-plane magnetization. For the three-dimensional spin-configuration of the domain wall, please refer to Supplementary Figure S2.

In panels A, B, the external field was oriented along the *x* and *y* direction, respectively, i.e. in the plane of the film. In these two cases, no sizeable displacement of the domain wall was observed following the application of the field. Only a slight deformation of the wall is observed when the field is applied along *x*. Importantly, the bottom panels show the absence of spin-wave emission from the wall when an oscillating field at 1.57 GHz was applied.

In panel C, the field was oriented along the *z* direction, i.e. out-of-plane. In this case, we observe a sizeable displacement of the wall according to the field direction. This behavior can be ascribed to the fact that the coupling of the field with the domain wall is mediated by the peculiar spin-configuration of the wall. In fact, the perfect compensation of the SAF system is broken in the out-of-plane direction in correspondence of the domain wall, which makes energetically favorable the emergence of an out-of-plane component of the magnetization with opposite sign at the two sides of the wall (see Supplmentary Note 3). It is worth noting that the coupling of the out-of-plane component of the RF field with the wall allows the efficient oscillation of the wall and the consequent SW emission even in case of perfectly compensated SAF. This is evident in the bottom panel, where spin-wave emission at both sides of the wall and propagation in the domains is observed. Noteworthy, in case of non-compensated SAF (i.e. SAF composed by layers of different materials or characterized by different thicknesses), the wall oscillation can be driven also via in-plane external field along *y*.

## Supplementary Note 5. Magnetic field generated by the stripline

The out of plane magnetic field necessary to excite the nanoscale spin-texture (see previous section) is provided by a stripline 5 μm wide patterned on top of the sample (see Methods). Figure S4A reports an optical image of the coplanar, which is aligned to the $Si_3N_4$ membrane in order to be in close proximity to the patterned domain wall, as shown in panel C.

The RF magnetic field generated by this structure can be evaluated by the Finite Element Method Magnetics simulation software. The parameters employed in both the simulation and the experiments are: 1.5 V peak-to-peak applied voltage, stripline thickness of 200 nm and DC resistance of 50 Ω. In panel B the out of plane (*z*) and the in plane (*y*) components of the RF magnetic field are plotted as a function of the distance from the edge of the stripline. The out of plane component of the field produced by the stripline is more than one order of magnitude higher than the in-plane one and it is maximum at the edge (about 1.4 mT), while it is slowly decaying outside the stripline, reaching a value of 0.25 mT after several μm.

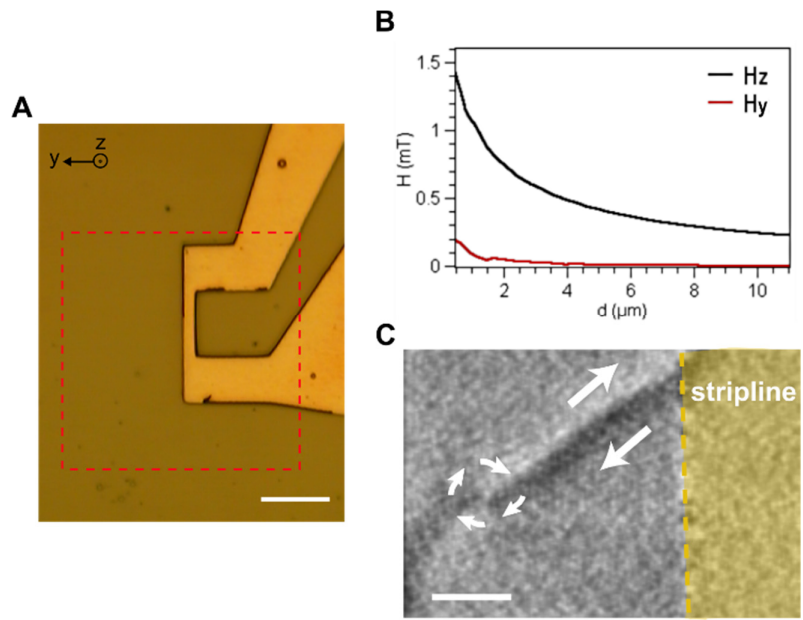

**Figure S4. Stripline characterization**. A. Optical image of the stripline patterned on top of the sample. The position of the underlying membrane is sketched in red, since it is not visible due to the poor optical contrast. Scale bar: 25 μm. B. FEMM simulation of the out of plane (*z*) and in plane (*y*) magnetic fields generated by the stripline as a function of the distance from its edge. C. XMCD static image of a straight domain wall with a vortex. White arrows indicate the equilibrium magnetization direction of the top layer and the chirality of the vortex. The edge of the stripline exciting the domain wall and the vortex is visible on the right, marked with a yellow dashed line. Scale bar: 1 μm

In Figure S4C, an XMCD static image of a straight domain wall and a vortex is shown (see Figure 3 of the main text). The edge of the stripline is visible and marked with a yellow dashed line. As mentioned in the main text, since the spin-wave emission is mediated mainly by the coupling between the out-of-plane field and the patterned spin textures, the domain wall can be placed at any angle with respect to the stripline. The out of plane field generated by the stripline is effective also in exciting spin waves emitted by more complex spin textures, such as the vortex (see Figures S4C) placed at about 3 μm from the edge.

## Supplementary Note 6. Generation of linear and radial wavefronts

Figure S5 shows the generation of planar (panel A) and radial (panel B) spin-wave wavefronts by straight domain walls and vortex-Bloch lines, respectively. In particular, panel A shows the STXM image of spin-waves excited by a straight domain wall, propagating away from the wall (see also Supplementary Movie 3). The black/white color corresponds to the oscillation of the out-of-plane component of the magnetization $M_z$ associated to the propagation of spin-waves, with respect to the average value across one cycle of excitation. The oscillation of the domain wall is driven by an external magnetic field $H_{RF}$, provided by a stripline run by radiofrequency current, patterned on top of the multilayer and in close proximity to the wall (see Methods). The magenta line indicates the domain wall and the white arrow indicates the direction of the magnetization in the top layer. The inset shows the spatial Fourier transform of the image. The high localization of the bright dots associated with the propagating mode demonstrates the highly directional emission from the wall.

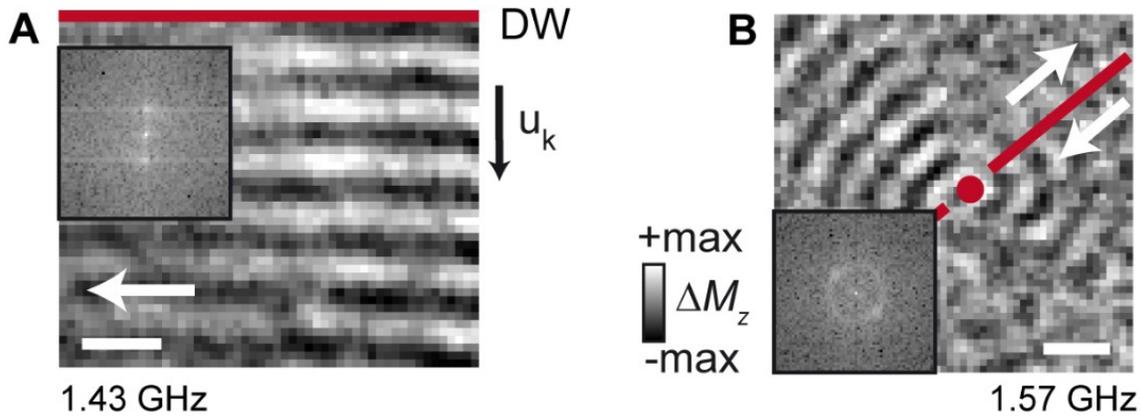

**Figure S5. Generation of linear and radial wavefronts**. A. Linear wavefronts generated by a straight wall. B. Radial wavefronts generated by a vortex Bloch line.

Panel B shows the excitation and propagation of spin-waves with radial symmetry, emitted from a vortex Bloch line located within a domain wall (see also Supplementary Movie 4). The radial symmetry of the spin-wave wavefronts is confirmed by the characteristic bright ring in the Fourier image in the inset. The brighter dots on the ring are related to the excitation of linear wavefronts from the domain wall. Noteworthy, the ring features a remarkably low ellipticity, which is the signature of the fact that the spin-wave wavelength is weakly dependent on the propagation direction (see detailed discussion in the main text).

## Supplementary Note 7. Additional micromagnetic simulations of spin-wave modes in SAF for Damon-Eshbach configuration ($\varphi = 90°$)

The extended antiferromagnetically coupled bilayer can support both acoustic and optic SW modes, depending on whether the magnetization of the two CoFeB films precess in-phase or out-of-phase, respectively.

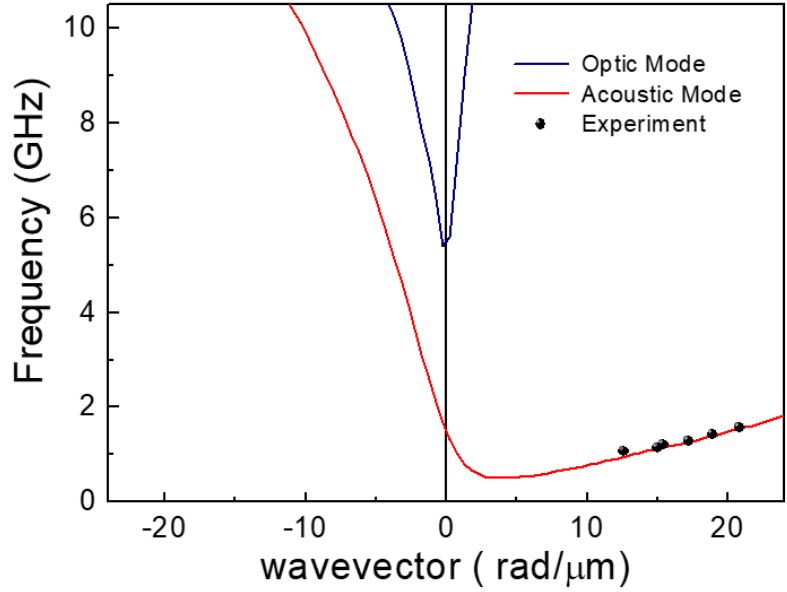

**Figure S6. Dispersion relation for spin waves in an extended bilayer system with antiparallel alignment of the CoFeB films**. Spin wave dispersion relation for acoustic (red line) and optic (blue line) modes. Filled black points represent the experimental dispersion.

Figure S6 shows the dispersion relation calculated at remanence for the configuration where the angle between *k* and the magnetization of the top CoFeB film is 90°. Micromagnetic simulations have been performed using the same parameters and the procedure described in the Methods. To put in evidence the presence of both the acoustic and optic modes, the magnetization dynamics has been excited using a sinc-shaped field pulse resulting from the superposition of a uniform spatial profile and one with a nodal plane through the film thickness. As it can be seen both the acoustic and the optic mode are present and are characterized by a non-reciprocal propagation.

Moreover, we performed additional simulations to calculate the modes spatial profiles at *k*=0. Micromagnetic simulations have been carried out using a simulated area having width 40 nm, length 40 nm, thickness 2×45 nm, and divided in 4×4×64 number of cells. Periodic boundary condition was applied both in *x* and *y* direction. The exchange bias field was modeled applying in the uppermost layer of cells of the top CoFeB film an external magnetic field of 30 mT parallel to the magnetization of the CoFeB film itself. The magnetization of the two CoFeB films was initialized to be pointing in opposite direction at remanence. The magnetization dynamics was excited applying in the entire area an external field sinc-pulse, $b(t) = b_0 \frac{sin(2\pi f_0(t-t_0))}{2\pi f_0(t-t_0)}$ oriented along the z-axis with an amplitude of $h_0$ = 10 mT and a frequency of $f_0$ = 30 GHz.

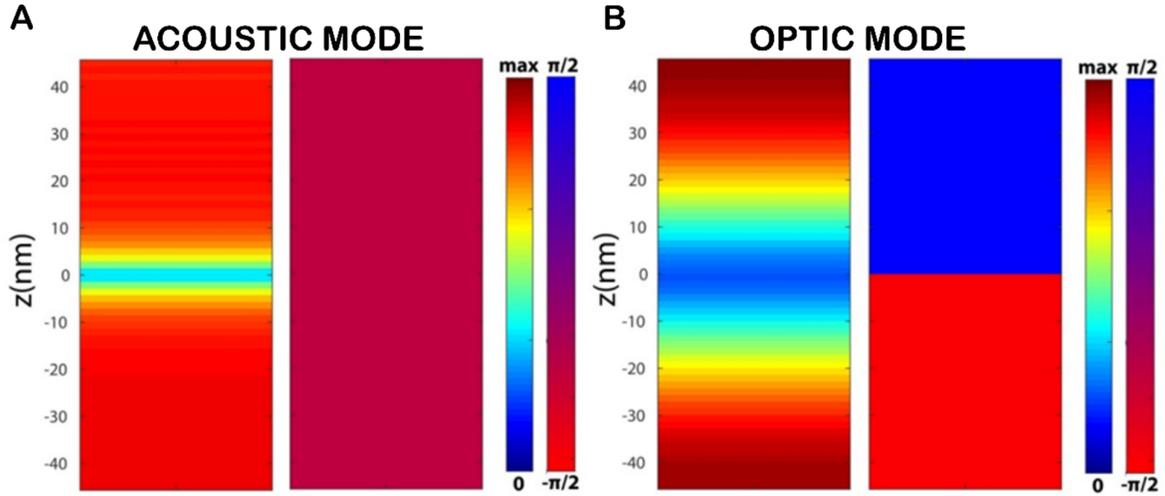

**Figure S7. Mode spatial profiles.** Amplitude (left panels) and phase (right panels) of the $M_z$ component of the dynamic magnetization of A the acoustic and B the optic modes calculated for $k = 0$, using a pulse with a uniform spatial profile and a pulse with a nodal plane through the film thickness, respectively.

Figure S7 reports the calculated amplitude (left panels) and the phase (right panel) of the $M_z$ component of the dynamic magnetization of the mode excited using a pulse with A a uniform spatial profile and B a pulse with a nodal plane through the film thickness. As it can be seen, depending on the pulse spatial profile, a mode characterized by an in-phase (out-of-phase) precession in the top and bottom CoFeB layers, is excited at about 1.5 (5.4) GHz. These results confirm the acoustic and optic character of the modes observed in the relation dispersion at lower and higher frequencies, respectively.

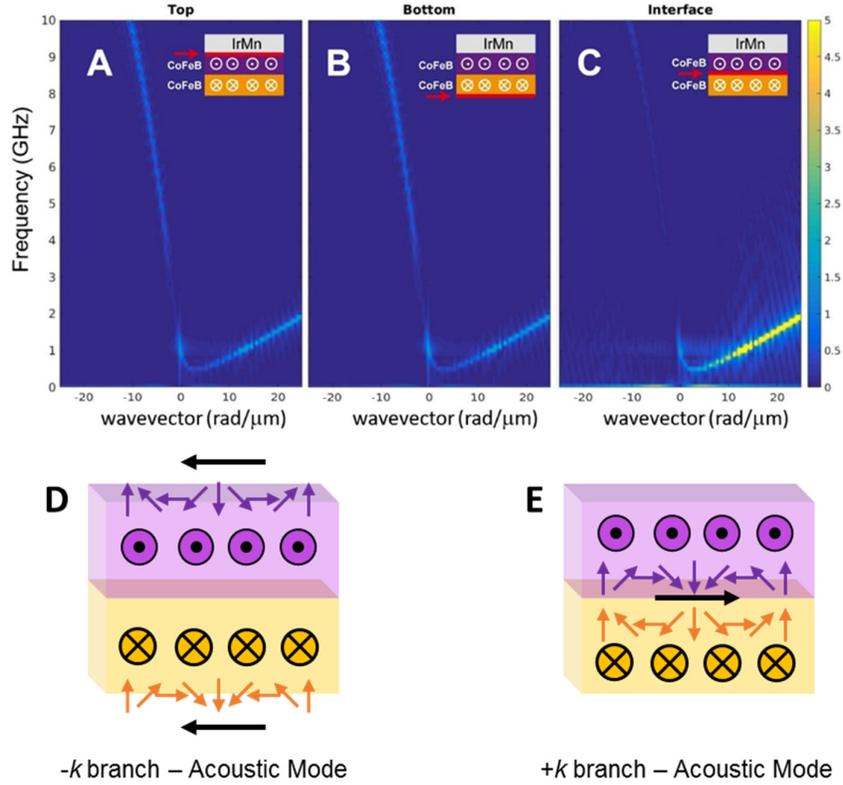

**Figure S8. Dispersion relation for the acoustic mode in an extended bilayer system with antiparallel alignment of the CoFeB films.** Spin wave dispersion relation for the acoustic mode. In panels A and B the dispersion is related to the modes in the uppermost layer of cells of the top CoFeB film, and in the lowermost layer of cells of the bottom CoFeB film, respectively. In panel C the dispersion refers to the modes in the layer of cells at the interface between the two CoFeB films. In panels (d) and (e), sketch of the modes belonging to the –$k$ and +$k$ branches, respectively.

Note that in our experiment only the acoustic mode has been observed, for the following reasons. First, the frequency of the optic mode is out of the frequency range experimentally investigated. Second, the total $M_z$ over the whole film is zero in the case of optic modes, therefore it would not be possible to observe it in our experimental STXM configuration.

As discussed in the main text, the acoustic mode is characterized by a strong non-reciprocal propagation, and as a consequence two branches, propagating along negative and positive $k$, respectively, can be identified. Moreover, we found that the two branches of the acoustic mode are characterized by a different spatial localization, in agreement with the theoretical calculations of Grünberg[2]. Figure S8 shows the simulated dispersion relation of the acoustic mode, calculated at remanence for the configuration where the angle between $k$ and the magnetization of the top CoFeB film is 90°, probing the dynamic component of the magnetization perpendicular to the film surface ($M_z$) in different regions of the sample. Micromagnetic simulations have been performed using the same parameters and the procedure described in the Methods. The dispersions reported in panels A and B have been obtained recording $M_z$ only in the uppermost layer of cells of the top CoFeB film, and in the lowermost layer of cells of the bottom CoFeB film, respectively. The dispersion showed in panel C has been achieved recording $M_z$ only in the layer of cells at the interface between the two CoFeB films. One can see that the mode propagating for positive $k$ has the maximum spin wave amplitude in the region at the interface between the two CoFeB films. On the contrary, the oscillation amplitude of the -$k$ mode is mainly concentrated at the top and bottom surfaces of the system. In panels D and E a sketch of the spin-waves modes belonging to the negative and positive $k$ branches, respectively, is reported.

**Supplementary Note 8. Simulations and STXM of spin-wave propagation towards the DW.**

In this section and related videos, we show STXM experiments and micromagnetic simulations of the spin-wave propagation in proximity of a domain wall, where the magnetization in the two SAF layers have opposite orientation with respect to the experiments shown in Figures 1-3 of the main manuscript. Figure S9A shows the spin-configuration in the SAF system in correspondence of the domain wall (in red), and the corresponding spin-wave wavevector $k_{SW}$ of the short-wavelength propagating mode, according to the dispersion relation shown in Figure 4A of the main manuscript. Figure S9B shows a frame of Supplementary Movie 7, where we show the result of micromagnetic simulations of short-wavelength spin-waves propagating towards the domain wall. For spin-wave excitation, an external oscillating monochromatic field of 0.2 mT at $f$ = 1.07 GHz oriented in the out-of-plane direction was applied to the whole system. As opposed to the configuration shown in the main manuscript, where short-wavelength spin-waves are emitted from the oscillating domain wall, here spin-waves travel towards the domain wall. Consistently with the spin-wave dispersion of Figure 4, no emission of spin-waves from the wall is observed in this configuration.

Supplementary Movie 8 shows STXM experiments showing short-wavelength spin-wave propagation towards the domain wall, in agreement with the simulations shown in Figure S9 and the spin-wave dispersion shown in Figure 4. The magnetization direction is consistent with an opposite orientation with respect to the configuration of Figures 1-3 of the main paper. In the STXM movies and simulations, the black/white contrast represents $\Delta M_z$.

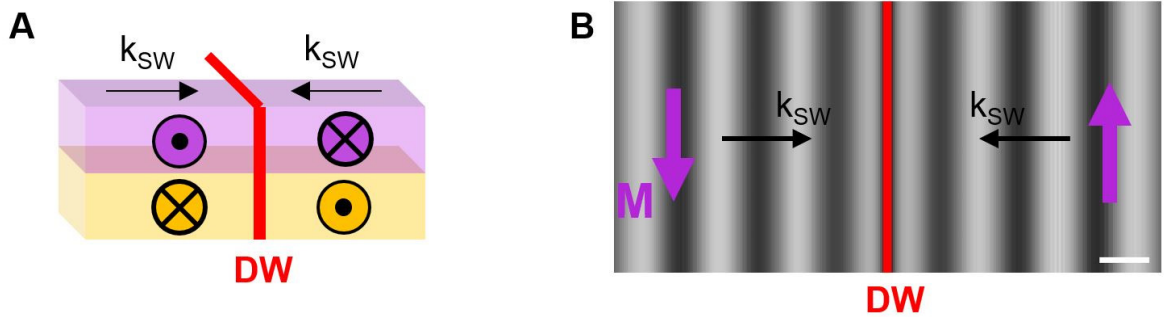

**Figure S9. Simulations of spin-wave propagation towards the domain wall.** A. Sketch of the magnetization configuration, showing the direction of the magnetization in proximity of a domain wall, in the two coupled ferromagnetic layers composing the SAF. B. Frame extracted from a micromagnetic simulation video showing spin-wave propagation towards the domain wall, for the magnetization configuration sketched in panel A at $f$ = 1.07 GHz.. The black/white contrast indicates the $\Delta M_z$ component of the magnetization. The spin-wave wavevector $k_{SW}$ and the static magnetization direction of the top layer $M$ are indicated. Scale bar 250 nm.

**Supplementary Note 9. Spin-wave backscattering by an isolated defect in a single layer film**

Here, we show the results of micromagnetic simulations of the emission and propagation of a spin-wave packet in the Damon-Eshbach configuration ($\varphi$ = 90°), in a uniformly magnetized CoFeB 5 nm monolayer slab. A 200 nm wide cylindrical defect is placed in the center of the monolayer, at a 2.5 µm distance (identical to the defect shown in Figure 4E of the main manuscript), for investigating the effect of a scattering center on the spin-wave propagation. In the top panel, spin-waves are emitted by a magnetic field line excitation along the $z$ (out-of-plane) direction (green line in Figure S10), 5 oscillation periods long, at 9 GHz. At this frequency, the spin-wave wavelength is comparable to the case of SAF shown in Figure 4E. Spin-waves are emitted on both sides with opposite $k$ vector. In the central panels, the interaction of the propagating packet with the defect is shown. In particular, the backscattered wave, indicated by the dashed line, has an amplitude which is comparable to the amplitude of the forward-propagating wave.

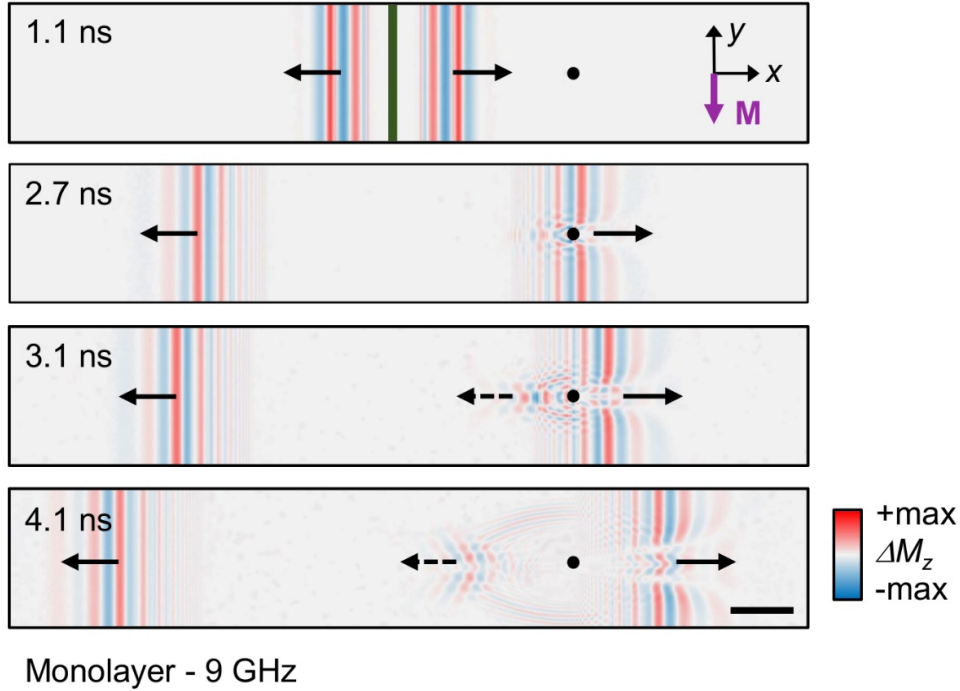

**Figure S10. Simulations of spin-wave backscattering by defects in a monolayer.** Simulations of spin-wave backscattering by a 200 nm diameter circular defect in a 5 nm CoFeB monolayer in the Damon-Eshbach configuration. Scale bar: 1 μm.

## Supplementary Note 10. Scanning transmission X-ray microscopy movies

The temporal evolution of the spin-wave emission and propagation can be better appreciated in the time-resolved STXM movies. Each frame of the movies is a time-resolved image acquired stroboscopically via STXM and shows the normalized $M_z$ contrast calculated as the magnetic deviation $\Delta M_z(t)$ from the time-averaged $<M_z(t)>$ state. The excitation of the spin-waves was applied through a sinusoidal function with frequency $f$. The time resolution of each movie was $\Delta t = 1/(7f)$, as reported in the Methods. The complete sequence corresponds to a period of sinusoidal excitation of the stripline.

| File name | Functionality | Freq. | Notes | Size (nm²) |
|---|---|---|---|---|
| Supplementary Movie 1 | Convex Wavefronts | 1.57 GHz | See Fig. 2a | 3800x5000 |
| Supplementary Movie 2 | Spin-wave Focusing | 1.43 GHz | See Fig. 2e | 4200x4600 |
| Supplementary Movie 3 | Linear Wavefronts | 1.43 GHz | See Fig. S3a | 2650x2250 |

| Supplementary Movie 4 | Radial Wavefronts | 1.57 GHz | See Fig. S3b | 2700x2700 |
| Supplementary Movie 5 | Interference 1 | 1.29 GHz | See Fig. 4a | 4500x3600 |
| Supplementary Movie 6 | Interference 2 | 1.57 GHz | See Fig. 4e | 3950x3900 |
| Supplementary Movie 7 | Towards DW simulations | 1.57 GHz | See Fig. S8 | 2560x1280 |
| Supplementary Movie 8 | Towards DW | 1.07 GHz | See Fig. S8 | 4000x2700 |